\documentclass[12pt,twoside]{article}

\evensidemargin=\oddsidemargin
\def\Title#1#2#3{%
    \baselineskip=18pt
    \begin{center}
          {\large\bf{#1} \\ }
          \bigskip\bigskip
          {#2} \\
          {#3} \\
    \end{center}}
\long\def\Abstract#1{%
         \bigskip
         \parbox{0.93\textwidth}{%
                 \begin{center}
                       {\bf Abstract} \\
                 \end{center}
                 \medskip{\baselineskip=14pt #1}
                 \vss}
         \bigskip}

\begin{document}

\Title{A view on the problems of Quantum Gravity}%
{T. P. Shestakova}%
{Department of Theoretical and Computational Physics,
Southern Federal University,\\
Sorge St. 5, Rostov-on-Don 344090, Russia \\
E-mail: {\tt shestakova@sfedu.ru}}

\Abstract{The existing approaches to quantization of gravity aim at giving quantum description of 3-geometry following to the ideas of the Wheeler -- DeWitt geometrodynamics. In this description the role of gauge gravitational degrees of freedom is missed. A probable alternative is to consider gravitational dynamics in extended phase space, taking into account the distinctions between General Relativity and other field theories. The formulation in extended phase space leads to some consequences at classical and quantum levels. At the classical level, it ensures that Hamiltonian dynamics is fully equivalent to Lagrangian dynamics, and the algebra of Poisson brackets is invariant under reparametrizations in a wide enough class including reparametrizations of gauge variables, meantime in the canonical Dirac approach the constraints' algebra is not invariant that creates problems with quantization. At the quantum level, the approach come to the description in which the observer can see various but complementary quantum gravitational phenomena in different reference frames that answers the spirit of General Relativity and Quantum Theory. Though until now the approach was applied to General Relativity in its original formulations, its implementation in different trends, including Quantum Loop Gravity or some other representations of gravitational variables, would also be of interest.}

\vspace{1cm}
I am grateful to the Organizers of the Conference for the opportunity to present my point of view on the problems of Quantum Gravity. In our attempts to quantize gravity we inevitably rely on approaches which work satisfactory for ordinary field theories, often without a careful analysis of applicability one or another approach to gravitation. An attractive point of Loop Quantum Gravity is that it aims at searching for its own way of description of Quantum Geometry and the Hilbert space of quantum states.

At the same time, Loop Quantum Gravity inherits the ideas of the Wheeler -- DeWitt Quantum Geometrodynamics which deals with 3-geometry being described by 3-metric $g_{ij}$, while the other components of metric tensor, $g_{0\mu}$, are missed in this consideration. They are traditionally believed to be redundant variables, whose role in General Relativity is just to fix a reference frame, and the choice of a reference frame should not affect physical phenomena. The purpose of this my talk is to demonstrate that, on the opposite, gauge variables $g_{0\mu}$ play an important role both at classical and quantum levels and must be taken into account in any quantization scheme.

The Wheeler -- DeWitt Quantum Geometrodynamics \cite{DeWitt} is based on three cornerstones: Dirac approach to quantization of systems with constraints \cite{Dirac1,Dirac2}, the Arnowitt -- Deser -- Misner (ADM) parametrization \cite{ADM} and the ideas of Wheeler that the wave functional must depend only on 3-geometry \cite{Wheeler1,Wheeler2}. In the Dirac approach the central part is given to a postulate, according to which each constraint $\varphi_m(q,p)=0$ after quantization becomes a condition on a state vector, or wave functional, $\Psi$:
$\varphi_m\Psi=0$. Let us emphasize that it is indeed a postulate, since it cannot be justified by any reference to the correspondence principle. As a consequence, in Quantum Geometrodynamics all the theory is reduced to the only Wheeler -- DeWitt equation, that, in its turn, leads to the problem of time and other related problems which cannot be resolved in the limits of Quantum Geometrodynamics.

The ADM parametrization enables one to write gravitational constraints in the form independent of gauge variables -- the lapse and shift functions $N$, $N_i$. It gave rise to an illusion that the theory in which the main equations are those of constraints does not depend on a choice of gauge conditions. However, it can be demonstrated that any choice of parametrization implies implicitly a certain choice of a reference frame. It is important to realize that, while at classical level the existing of constraints reflects gauge invariance of the theory, at quantum level imposing operator form of constraints on the wave functional does not ensure gauge invariance. Strictly speaking, we do not have any proof that the Wheeler -- DeWitt theory is gauge-invariant. In this situation the quantum Hamiltonian constraint loses its sense and the whole procedure of quantization becomes questionable.

In our attempts to reconcile General Relativity with quantum principles we must be sure that we are quantizing General Relativity, and not any other theory which resembles General Relativity by its structure but is not equivalent to it. Several years ago a statement appeared in literature \cite{KK}, that the ADM formulation \cite{ADM} is not equivalent to the Dirac formulation of Hamiltonian dynamics for gravity \cite{Dirac3}, because the both formulations are not related by a canonical transformation. The raised question is of fundamental nature, since, if the statement were correct, we would have to agree with the declaration that ``the ADM formulation might be considered as a model (geometrodynamics or ADM gravity) without any reference to Einstein GR'' (\cite{KK}, p. 62). In fact, this statement is a result of misunderstanding. The transformation from components of metric tensor to the ADM variables,
\begin{equation}
\label{ADM-tr}
g_{00}=\gamma_{ij}N^i N^j-N^2,\qquad
g_{0i}=\gamma_{ij}N^j,\qquad
g_{ij}=\gamma_{ij},
\end{equation}
concerns gauge degrees of freedom which, from the viewpoint of Dirac, are not canonical variables at all. To pose the question, if the transformation (\ref{ADM-tr}) is canonical, one should extend the original phase space including into it gauge degrees of freedom and their momenta and also introduce into the Lagrangian missing velocities corresponding to gauge variables. It can be done by means of special (differential) gauge conditions that {\it actually extends} the phase space. For general enough gauge conditions, $f^{\mu }(g_{\nu\lambda})=0$, their differential form reads
\begin{equation}
\label{diff-g}
\frac{d}{dt}f^{\mu}(g_{\nu\lambda})=0,\qquad
\frac{\partial f^{\mu}}{\partial g_{00}}\dot g_{00}
 +2\frac{\partial f^{\mu}}{\partial g_{0i}}\dot g_{0i}
 +\frac{\partial f^{\mu}}{\partial g_{ij}}\dot g_{ij}=0.
\end{equation}
Then, we can write the effective action including gauge and ghost sectors as it appears in the path integral approach to gauge field theories,
\begin{equation}
\label{full-act}
S=\int d^4 x\left({\cal L}_{(grav)}+{\cal L}_{(gauge)}+{\cal L}_{(ghost)}\right).
\end{equation}
It has been shown \cite{Shest1} for the full gravitational theory that the transformation
\begin{equation}
\label{new-var}
g_{0\mu}=v_{\mu}\left(N_{\nu},g_{ij}\right),\qquad
g_{ij}=\gamma_{ij},
\end{equation}
where $v_{\mu}\left(N_{\nu},g_{ij}\right)$ is invertible function, is canonical in extended phase space embracing spatial components of metric tensor $g_{ij}$ which are canonical in Dirac's sense, as well as gauge variables $g_{0\mu}$, ghosts and their momenta. The transformation to the ADM variables (\ref{ADM-tr}) is a particular case of (\ref{new-var}). The canonicity of the transformation (\ref{new-var}) not just restores the status of the ADM formulation as completely equivalent to the original Einstein formulation, but also demonstrates the role of gauge degrees of freedom which are responsible for consistency of the theory. By excluding gauge degrees of freedom from consideration, we would reduce the full gravitational theory to its particular case answering to a particular choice of parametrization and gauge conditions when the possibility to go over to other parametrizations being missed. From this point of view we can consider the Wheeler -- DeWitt geometrodynamics as an attempt to quantize gravity in the ADM parametrization and under particular gauge conditions $N=1$, $N_i=0$ \cite{DeWitt}.

The extension of phase space by means of introducing a differential form of gauge conditions was suggested in the works \cite{SSV1,SSV2,SSV3,SSV4}. It is well known that the idea of extended phase space was put forward by Batalin, Fradkin and Vilkovisky (BFV) \cite{BFV1,BFV2,BFV3}. However, in their approach gauge variables were still considered as non-physical, secondary degrees of freedom playing just an auxiliary role in the theory. While the BFV approach aimed at reproducing the results of Dirac's canonical quantization on a path integral level, our construction of extended phase space guarantees equivalence of Lagrangian and Hamiltonian dynamics of a constrained system for a wide enough class of parametrizations, and constraints and gauge condition get a status of Hamiltonian equations in extended phase space \cite{Shest1}. The algebra of Poisson brackets turns out to be invariant under reparametrizations from this class.

Based upon this formulation of Hamiltonian dynamics, we have proposed Quantum Geometrodynamics in extended phase space \cite{SSV3,SSV4} making use of the path integration approach which is more powerful. We paid a special attention to the peculiarities of General Relativity. In ordinary quantum theory its gauge invariance is ensured by asymptotic boundary conditions in the path integral. The boundary conditions imply the existing of asymptotic states in which gauge and physical degrees of freedom could be separated from each other. In the case of gravity asymptotic states exists only in asymptotically flat spacetime, but for a non-trivial spacetime topology imposing asymptotic boundary conditions is not justified. Thus, path integral quantization as well as canonical quantization do not enable us to prove gauge invariance of the theory.

In this situation the requirement for the wave function to satisfy the Wheeler -- DeWitt equation seems to be too strong and not well-grounded. However, independently on our notion about gauge invariance or noninvariance of the theory, we expect that the wave function has to obey some Schr\"odinger equation. One can derive the Schr\"odinger equation from the path integral by the well-known standard procedure and only after that one can investigate the question if there exist any conditions which would ensure gauge invariance. I believe that this way is more  logically consistent.

Since asymptotic boundary conditions are not justified in gravitational theory, one should define the wave function on extended configurational space depending on gauge and physical gravitational variables and ghosts (one may also add matter fields). So, the wave function is assumed to be a solution to the equation
\begin{equation}
\label{full_Schr}
H\Psi\left[g_{0\mu},\,g_{ij},\,\theta^{\nu},\,\bar\theta_{\nu},\,\phi\right]
 =i\frac{\partial\Psi}{\partial t}.
\end{equation}
$H$ is the Hamiltonian in extended phase space. The problem of time does not appear is this approach though the physical sense of time in Quantum Gravity require additional clarification. In any case, the existing of time is related with a reference frame in which we study spacetime geometry.

The structure of the wave function is determined by the structure of the path integral with the effective action (\ref{full-act}). It has been demonstrated for models with finite numbers degrees of freedom that the part of the wave function depending on gauge and ghost variables can be separated out. For example, for the Bianchi IX model one gets:
\begin{equation}
\label{GS-A}
\Psi(\tilde N,\,Q,\,\theta,\,\bar\theta;\,t)
 =\int\Psi_{(ph)}(Q,\,t)\,\delta(\tilde N-f(Q)-k)\,(\bar\theta+i\theta)\,dk.
\end{equation}
$\tilde N$ stands for an arbitrary gauge variable, and $Q$ stands for a set of physical variables. The physical part of the wave function satisfies the equation
\begin{equation}
\label{phys.SE}
H_{(ph)}\Psi_{(ph)}(Q,\,t)
 =i\,\frac{\partial\Psi_{(ph)}(Q,\,t)}{\partial t}
\end{equation}
The physical Hamiltonian $H_{ph}$ depends, in general, on a chosen parametrization and gauge. In particular, for the ADM parametrization and the condition $N=1$ the left-hand side of (\ref{phys.SE}) coincides with the left-hand side of the Wheeler -- DeWitt equation \cite{Shest2}. It gives another evidence that the Wheeler -- DeWitt Geometrodynamics is not a gauge invariant theory.

In the proposed version of Quantum Geometrodynamics spacetime geometry is described from the viewpoint of the observer who can see various but complementary quantum gravitational phenomena in different reference frames. It contradicts to the original goal to construct gauge-invariant Quantum Gravity, however, to a greater extent it answers the spirit of General Relativity and Quantum Theory where the observer plays the central role in description of physical phenomena.

I believe that it would be rather instructive to apply some ideas of extended phase space approach to Loop Quantum Gravity, or, at least, loop quantum cosmological models. The first step of the program would be the formulation of dynamics of General Relativity in new variables, in such a way that this formulation would be equivalent to the original one. It would ensure that the theory have the correct classical limits. And this is the point where the extended phase space approach could play its role, since it allows us to work with different parametrizations. On the next, quantum stage one could explore without prejudice the question if Quantum Gravity can be constructed in a gauge-invariant way. I hope that this study would enrich our understanding of quantization of Gravity.

\section*{Acknowledgements}
My participation in the Loops 11 conference was partially supported by the RFBR grant 11-02-08088.


\small
\begin{thebibliography}{99}
\itemsep=-5pt
\bibitem{DeWitt}
B. S. DeWitt,
 {\it Phys. Rev.\/} {\bf 160} (1967), P. 1113--1148.
\bibitem{Dirac1}
P. A. M. Dirac,
 {\it Can. J. Math.\/} {\bf 2} (1950), P. 129--148.
\bibitem{Dirac2}
P. A. M. Dirac,
 {\it Proc. Roy. Soc.\/} {\bf A246} (1958), P. 326--332.
\bibitem{ADM}
R. Arnowitt, S. Deser and C. W. Misner,
 in: {\it Gravitation, an Introduction to Current Research\/},
 ed. by L. Witten, John Wiley \& Sons, New York (1962) P. 227--265.
\bibitem{Wheeler1}
J. A. Wheeler,
 in: {\it Relativity, Groups and Topology\/},
 eds. C. DeWitt, B. S. DeWitt, Gordon \& Breach, New York (1964).
\bibitem{Wheeler2}
J. A. Wheeler,
 {\it Einstein's vision\/}, Springer Verlag, Berlin--Heidelberg--New York (1968).
\bibitem{KK}
N. Kiriushcheva and S. V. Kuzmin,
 {\it Central Eur. J. Phys.\/} {\bf 9} (2011), P. 576--615.
\bibitem{Dirac3}
P. A. M. Dirac,
 {\it Proc. Roy. Soc.\/} {\bf A246} (1958), P. 333--343.
\bibitem{Shest1}
T. P. Shestakova,
 {\it Class. Quantum Grav.\/} {\bf 28} (2011), 055009.
\bibitem{SSV1}
V. A. Savchenko, T. P. Shestakova and G. M. Vereshkov,
 {\it Int. J. Mod. Phys.\/} {\bf A14} (1999), P. 4473--4490.
\bibitem{SSV2}
V. A. Savchenko, T. P. Shestakova and G. M. Vereshkov,
 {\it Int. J. Mod. Phys.\/} {\bf A15} (2000), P. 3207--3220.
\bibitem{SSV3}
V. A. Savchenko, T. P. Shestakova and G. M. Vereshkov,
 {\it Gravitation \& Cosmology\/} {\bf 7} (2001), P. 18--28.
\bibitem{SSV4}
V. A. Savchenko, T. P. Shestakova and G. M. Vereshkov,
 {\it Gravitation \& Cosmology\/} {\bf 7} (2001), P. 102--116.
\bibitem{BFV1}
E. S. Fradkin and G. A. Vilkovisky,
 {\it Phys. Lett\/} {\bf B55} (1975), P. 224--226.
\bibitem{BFV2}
I. A. Batalin and G. A. Vilkovisky,
 {\it Phys. Lett\/} {\bf B69} (1977), P. 309--312.
\bibitem{BFV3}
E. S. Fradkin and T. E. Fradkina,
 {\it Phys. Lett\/} {\bf B72} (1978), P. 343--348.
\bibitem{Shest2}
T. P. Shestakova,
 {\it Gravitation \& Cosmology\/} {\bf 12} (2006), P. 223--226.
\end{thebibliography}
\end{document}